\RequirePackage{lineno} 
\documentclass[twocolumn,showpacs,aps,prl,superscriptaddress,linenumbers,letterpaper]{revtex4}

\usepackage{hyperref}
\usepackage[hyphenbreaks]{breakurl}
\usepackage{graphicx}
\usepackage{dcolumn}
\usepackage{amsmath}
\usepackage{epsfig}
\usepackage{xcolor}

\input babarsym

\newcommand{\BABARPubYear}    {14}
\newcommand{\BABARPubNumber}  {010}

\newcommand{\SLACPubNumber} {16226}

\def\mee{\ensuremath{m_{\epem}}}
\def\mmumu{\ensuremath{m_{\mumu}}}
\def\memu{\ensuremath{m_{e^\pm\mu^\mp}}}
\def\mpipi{\ensuremath{m_{\pi^+\pi^-}}}
\def\mKK{\ensuremath{m_{K^+K^-}}}
\def\mKpi{\ensuremath{m_{K^\pm\pi^\mp}}}
\def\pt{\ensuremath{p_T}}
\def\mc{\ensuremath{\rm MC}}

\begin{document}

\preprint{\babar-PUB-\BABARPubYear/\BABARPubNumber} 
\preprint{SLAC-PUB-\SLACPubNumber} 

\begin{flushleft}
\babar-PUB-\BABARPubYear/\BABARPubNumber\\
SLAC-PUB-\SLACPubNumber\\[10mm]
\end{flushleft}

\title{{\large \bf
Search for Long-Lived Particles in \boldmath{$e^+e^-$} Collisions
} }

%
\author{J.~P.~Lees}
\author{V.~Poireau}
\author{V.~Tisserand}
\affiliation{Laboratoire d'Annecy-le-Vieux de Physique des Particules (LAPP), Universit\'e de Savoie, CNRS/IN2P3,  F-74941 Annecy-Le-Vieux, France}
\author{E.~Grauges}
\affiliation{Universitat de Barcelona, Facultat de Fisica, Departament ECM, E-08028 Barcelona, Spain }
\author{A.~Palano$^{ab}$ }
\affiliation{INFN Sezione di Bari$^{a}$; Dipartimento di Fisica, Universit\`a di Bari$^{b}$, I-70126 Bari, Italy }
\author{G.~Eigen}
\author{B.~Stugu}
\affiliation{University of Bergen, Institute of Physics, N-5007 Bergen, Norway }
\author{D.~N.~Brown}
\author{L.~T.~Kerth}
\author{Yu.~G.~Kolomensky}
\author{M.~J.~Lee}
\author{G.~Lynch}
\affiliation{Lawrence Berkeley National Laboratory and University of California, Berkeley, California 94720, USA }
% ...
\collaboration{The \babar\ Collaboration}
\noaffiliation

\begin{abstract}

We present a search for a neutral, long-lived particle $L$ that is
produced in $\epem$ collisions and decays at a significant distance
from the $\epem$ interaction point into various flavor combinations of
two oppositely charged tracks.  The analysis uses an $\epem$ data
sample with a luminosity of $489.1 \invfb$ collected by the
\babar\ detector at the $\FourS$, $\ThreeS$, and $\TwoS$ resonances and just
below the $\FourS$. Fitting the two-track mass distribution in search of
a signal peak, we do not observe a significant signal, and set 90\%
confidence level upper limits on the product of the $L$ production
cross section, branching fraction, and reconstruction efficiency for
six possible two-body $L$ decay modes as a function of the $L$ mass.
The efficiency is given for each final state as a function of the
mass, lifetime, and transverse momentum of the candidate, allowing
application of the upper limits to any production model.  In addition,
upper limits are provided on the branching fraction $\BR(B\to X_s L)$,
where $X_s$ is a strange hadronic system.
\end{abstract}


\maketitle

Recent anomalous astrophysical
observations~\cite{ref:anomalies1,ref:anomalies2,ref:anomalies3}
have generated interest in GeV-scale hidden-sector states that may
be long-lived~\cite{ref:pospelov,ref:essig,ref:fabio,ref:yavin,ref:inflaton,ref:susy1,ref:susy2,ref:higgsP,ref:nelson}.
Searches for long-lived particles have been performed both in the
sub-GeV~\cite{ref:limitSummary,ref:limitSummary2,ref:nuTev} and
multi-GeV~\cite{ref:d00,ref:d0,ref:cdf,ref:atlas1,ref:atlas2,ref:atlasLLP}
mass ranges.  Dedicated experiments to search for such particles have
been proposed~\cite{ref:gninenko} or are under
construction~\cite{ref:heavyPhoton}.  However, the $O(1\gevcc)$ mass
range has remained mostly unexplored, especially in a heavy-flavor
environment.  $B$~factories offer an ideal laboratory to probe this
regime. Previously, the only $B$-factory results were from a search
for a heavy neutralino by the Belle
collaboration~\cite{Liventsev:2013zz}.

We search herein for a neutral, long-lived particle $L$, which
decays into any of the final states $f = e^+e^-$, $\mu^+\mu^-$,
$e^{\pm}\mu^{\mp}$, $\pi^+\pi^-$, $K^+K^-$, or $K^{\pm}\pi^{\mp}$.
A displaced vertex and two-body decay kinematics constitute
the main means for background suppression, and the search is performed
by fitting the distribution of the $L$-candidate mass.

The results are presented in two ways.  In the ``model-independent''
presentation, no assumption is made regarding the production mechanism
of the $L$. Rather, we present limits on the product of the inclusive
production cross section $\sigma(\epem\to LX)$, branching fraction
$\BR(L\to f)$, and efficiency $\epsilon(f)$ for each of
the two-body final states $f$, where $X$ is any set of particles. As
supplemental material to this letter~\cite{ref:supp}, we provide
tables of the efficiency as a function of $L$ mass $m$,
transverse~\footnote{The term ``transverse'' refers throughout this
  paper to projections of vectors onto the plane transverse to the
  direction of the $\epem$ center of mass system.}  momentum $\pt$ in
the center-of-mass (CM) frame, and proper decay distance $c\tau$,
assuming the $L$ to be a spin-zero particle.
The provided upper limits, efficiencies, and \pt\ distributions of the
simulated events used to obtain the efficiencies facilitate the
application of the model-independent presentation of the results to
any specific model of $L$ production.
In the ``model-dependent'' presentation, we provide limits on the
branching fraction for the decay $B\to X_s L$, where $X_s$ is a
hadronic system with strangeness $-1$. This presentation is motivated
by Higgs-portal models of dark matter and other hidden
sectors~\cite{ref:inflaton,ref:susy1,ref:susy2,ref:higgsP}.

The data were collected with the \babar\ detector at the PEP-II
asymmetric-energy \epem collider at SLAC National Accelerator
Laboratory. 
The sample consists of $404.0 \pm 1.7\invfb$
collected at a CM energy corresponding to the $\FourS$
resonance,
an ``off-resonance'' sample of $43.74 \pm 0.20\invfb$ collected about
$40\mev$ below the $\FourS$ peak,
$27.85 \pm 0.18\invfb$ collected at the $\ThreeS$, 
and $13.56 \pm 0.09\invfb$ taken at the $\TwoS$~\cite{luminosity}.
The $\FourS$ sample contains $(448.4 \pm 2.2)\times 10^6$ \BB pairs,
and the $\ThreeS$ and $\TwoS$ samples have 
$(121.3 \pm 1.2)\times 10^6$ $\ThreeS$ and
$(98.3 \pm 0.9)\times 10^6$ $\TwoS$ mesons, respectively~\cite{ref:Bcounting}.
An additional \FourS sample of $20.37 \pm 0.09 \invfb$ is used
to validate the analysis procedure and is not included in the final analysis.

The \babar\ detector and its operation are described in detail in
Refs.~\cite{babar} and~\cite{babar2}.  Charged-particle momenta are
measured in a tracking system consisting of a five-layer, double-sided
silicon vertex detector (SVT) and a 40-layer drift chamber (DCH), both
located in a 1.5~T axial magnetic field.  Electron and photon
energies are measured in a CsI(Tl) electromagnetic calorimeter (EMC) inside
the magnet coil.  Charged-particle identification (PID) is performed
using an internally reflecting, ring-imaging Cherenkov detector, as well as
the energy loss measured by the SVT, DCH and EMC.  Muons are identified
mainly with the instrumented magnetic-flux return.

Using Monte Carlo (MC) simulations, we determine both the signal mass
resolution and reconstruction efficiency.  The events are produced
with the \evtgen~\cite{ref:evtgen} event generator, taking the $L$
spin to be zero.
We generate two types of signal MC samples. 
In the first type, which is used to create the efficiency
tables~\cite{ref:supp} for the model-independent presentation, the $L$
is produced at 11 different masses, $m_0^{\mc}=0.5$, 1, 2, 3, 4,
5, 6, 7, 8, 9, and 9.5\gevcc.  
For $m_0^{\mc}\le 4\gevcc$, the $L$ is created in the process
$\epem\to \BB$, with one $B$ meson decaying to $L + N\pi$
($N=1$, 2, or 3) and the other $B$ decaying generically.  At higher
masses, the production process is $\FourS \to L + N\pi$.
In both cases, the $L$ is produced uniformly throughout the available
phase space, with an average transverse decay distance of $20\cm$.
The events are subsequently reweighted to obtain efficiencies for other
decay lengths. 
Note that these specific processes do not reflect preferred hypotheses
about the production mechanism, nor do the results depend on these
processes. Rather, they are a convenient method to populate the 
kinematic range for the efficiency tables.

The second type of signal MC, used for the model-dependent
presentation of the results, contains $B\to X_s L$ decays, for the seven 
mass values $m_0^{\mc}=0.5$, 1, 2, 3, 3.5, 4, and 4.5\gevcc.  The
$X_s$ is nominally taken to be 10\% $K$, 25\% $K^*(892)$, and 65\%
$K^*(1680)$~\cite{ref:pdg}, with the high-mass tail of the $X_s$
spectrum suppressed by phase-space limitations, especially for heavy
$L$ states. This choice of $X_s$ composition results in an
$L$-momentum spectrum as a function of $m_0^{\mc}$ that reproduces the dimuon
spectrum for $B\to X_s \mu^+\mu^-$ in events generated with \evtgen\
using the BTOXSLL model~\cite{ref:evtgen}.  The other $B$ meson in the
event decays generically.

In addition to the signal MC samples, background MC samples are used for
optimizing the event selection criteria and studying the
signal extraction method. The background samples are $\epem\to \BB$
(produced with \evtgen~\cite{ref:evtgen}), $\tautau$, $\mumu$ ({\tt
  KK2f}~\cite{ref:KK}), $\epem$ ({\tt BHWIDE}~\cite{ref:bhwide}), and
$\qqbar$ events ({\tt JETSET}~\cite{ref:jetset}), where $q$ is a $u$,
$d$, $s$, or $c$ quark.  The detector response is simulated with {\tt
  GEANT4}~\cite{geant4}.

The $L$ candidates are reconstructed from pairs of oppositely charged
tracks, identified as either $e^+e^-$, $\mu^+\mu^-$, $e^\pm \mu^\mp$,
$\pi^+\pi^-$, $K^+K^-$, or $K^\pm\pi^\mp$.
The PID efficiency depends on the track momentum,
and is in the range $0.96-0.99$ for electrons, $0.60-0.88$ for muons,
and $0.90-0.98$ for kaons and pions. The pion misidentification
probability is less than $0.01$ for the electron PID criteria, less than
$0.03$ for the muon criteria, and averages at $0.06$ for the kaon
criteria.
A track may have different PID assignments and may appear in multiple pairs.
Each track must satisfy $d_0 /\sigma_{d_0} > 3$, where $d_0$ is the
transverse distance of closest approach of the track to the
\epem\ interaction point (IP), and $\sigma_{d_0}$ is the $d_0$
uncertainty, calculated from the SVT and DCH hit position
uncertainties during the track reconstruction.
The two tracks are fit to a common vertex, and the $\chi^2$ value of the fit
is required to be smaller than 10 for one degree of freedom.
The two-dimensional vector $\vec r$ between the IP and the vertex in
the transverse plane must have length $r \equiv |\vec r|$ in the range
$1<r<50\cm$, and the uncertainty on $r$ is required to satisfy
$\sigma_r < 0.2\cm$.
We require the angle $\alpha$ between $\vec r$ and the $L$-candidate
transverse-momentum vector to satisfy $\alpha < 0.01$~rad.
The uncertainty $\sigma_m$ on the measured $L$-candidate mass $m$
must be less than $0.2\gevcc$.
The $L$ candidate is discarded if either of the tracks has SVT or DCH
hits located between the IP and the vertex, or if the vertex is within
the material of the beampipe wall, the DCH support tube, or the DCH
inner cylinder.
Candidates must satisfy the following decay-mode-specific
invariant-mass criteria: $\mee > 0.44\gevcc$, $\mmumu < 0.37\gevcc$ or
$\mmumu > 0.5\gevcc$, $\memu > 0.48\gevcc$, $\mpipi > 0.86\gevcc$,
$\mKK > 1.35\gevcc$, and $\mKpi > 1.05\gevcc$.  These criteria reject
background from $\KS \rightarrow \pi^+\pi^-$ and $\Lambda \rightarrow
p\pi^-$ decays.  In addition, other than in the \mumu mode, they
exclude low-mass regions in which the mass distributions of background
MC events are not smooth, and therefore incompatible with the
background description method outlined below.
We require at least one of the tracks of $L\to\mumu$ candidates with
$m \ge 8\gevcc$ to have an SVT hit.  This rejects candidates that
decay into $\mumu$ within the material of the final-focusing magnets, and thus have
poor mass resolution. 
These selection criteria are found to yield near-optimal signal
sensitivity given the broad range of $m$ and $r$ values of this
search.

For each decay mode, we determine the full efficiency $\epsilon$, including the
impact of detector acceptance, trigger, reconstruction, and selection
criteria, for different values of $m_0^{\mc}$, $c\tau$, and \pt. 
The efficiency, which is tabulated in Ref.~\cite{ref:supp}, reaches a
maximal value of $\epsilon=52\%$ for $L\to\pi^+\pi^-$ decays with $m =
2\gevcc$, $\pt >4\gevc$, and $c\tau= 6\cm$. The dominant factor
affecting $\epsilon$ is the average transverse flight distance
$\left< r\right> = c\tau\left<\pt\right>/m$.  Reflecting the
$1<r<50\cm$ requirement, $\epsilon$ drops rapidly when $\left<
  r\right>$ goes below  $1\cm$ or above $50\cm$. 
In addition, $\epsilon$ has some dependence on the $L$ polar-angle
$\theta$, measured with respect to the direction of the $\epem$ center
of mass.  For a $1+\cos^2\theta$ distribution in the CM frame, the
strongest dependence is observed for track momentum $p<0.3\gevc$,
where $\epsilon$  is decreased by 22\% relative to that of a
uniform $\cos\theta$ distribution.  For $p>2$~\gevc, $\epsilon$
varies by no more than 8\%.
Similarly, the efficiency depends weakly on whether $L$ is a scalar or
a vector particle. For a longitudinally polarized vector, $\epsilon$
typically varies by a few percent relative to the scalar
case, with the greatest impact being an efficiency reduction of 25\%
for $\pt<0.3\gevc$, $m=7\gevcc$.

The dominant source of background consists of hadronic events with high track
multiplicity, where large-$d_0$ tracks originate mostly from $\KS$,
$\Lambda$, $K^\pm$, and $\pi^\pm$ decays, as well as particle
interactions with detector material.
Random overlaps of such tracks comprise the majority of the background candidates.

We extract the signal yield for each final state as a function of $L$ mass
with unbinned extended maximum-likelihood fits of the $m$
distribution.  The procedure is based on the fact that signal MC events
peak in $m$ while the background distribution varies slowly.  The fit
probability density functions (PDFs) for signal and background are
constructed separately for each mode and each data sample.
The PDFs account for the signal mass resolution, which is
evaluated separately in each of 11 mass regions, where each region
straddles the $m_0^{\mc}$ value of one of the signal MC samples of the
first type.  
In region $i$, the value of the signal PDF for a candidate with hypothesis 
mass $m_0$, measured mass $m$, and mass resolution uncertainty $\sigma_{m}$ is
$P^i_S(m)=H^i_S\left((m-m_0)/\sigma_{m}\right)$, where $H^i_S(x)$ is the
histogram of the mass pull $x=(m^{\mc}-m_0^{\mc})/\sigma_m^{\mc}$ for
signal MC events of true mass $m_0^{\mc}$, measured mass $m^{\mc}$,
and $m^{\mc}$ uncertainty $\sigma_{m}^{\mc}$.
This PDF accounts correctly for the large variation in $\sigma_m$ 
with $r$ and $m$.

\begin{figure}[!htbp]
\includegraphics[width=\columnwidth, bb=40 20 550 670]{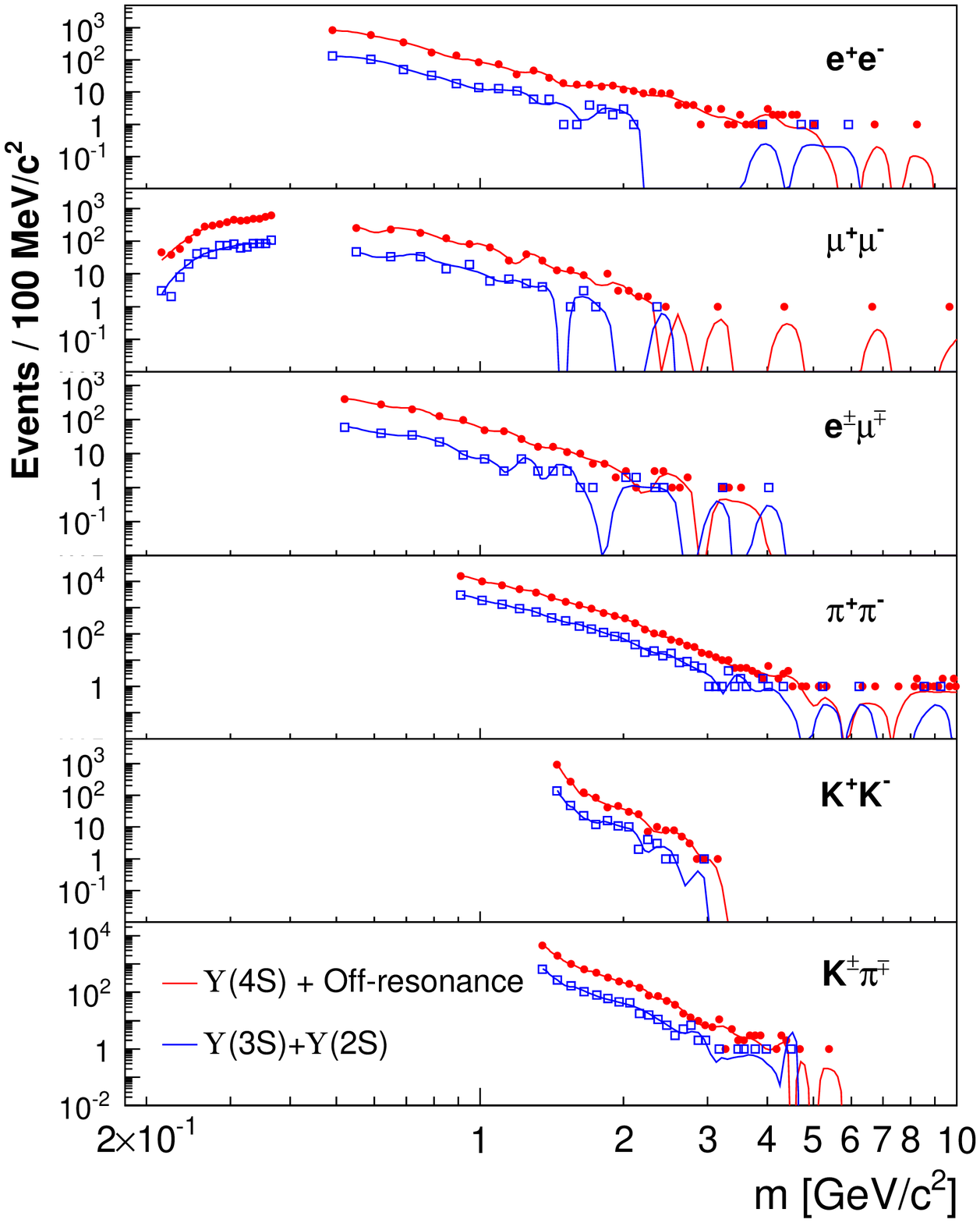}
\caption{\label{fig:res} Mass distribution of the $\FourS+{\rm
    off\mbox{-}resonance}$ data (red solid points) and
    $\Upsilon(3S)+\Upsilon(2S)$ data (blue open squares) for each
    mode, overlaid with the background PDF $P_B$ in matching color.
    In the \mumu mode, the bin width in the range $m < 370\mevcc$ is
    10\mevcc.}
\end{figure}

The background PDF $P_B(m)$ is obtained from the data, so as not to
rely on the background simulation, with the following procedure. First,
we create a variable-bin-width histogram $H_D(m)$ of the data $m$
distribution. The width of a histogram bin, whose lower edge is in $m$
region $i$, is $w_i = n R_i$, where $n = 15$, and
$R_i$ is the RMS width of the signal $m - m_0^{\mc}$ distribution in
that region. The value of $R_i$ ranges from about $6\mevcc$ for
$m_0^{\mc} = 0.5\gevcc$ to $180\mevcc$ for $m_0^{\mc} = 9.5\gevcc$.
We obtain $P_B(m)$ by fitting $H_D(m)$ with a second-order polynomial
spline, with knots located at the bin boundaries.
Simulation studies of the background mass distribution show that the
choice $n=15$ is sufficiently large to prevent $P_B(m)$ from conforming to
signal peaks and thus hiding statistically significant signals, yet
sufficiently small to avoid high false-signal yields due to background
fluctuations.
Fig.~\ref{fig:res} shows the $m$ distributions of the data (with uniform mass
bins) and the background PDFs.

We scan the data in search of an $L$ signal, varying $m_0$ in steps of
$2\mevcc$.  At each scan point, we fit the data in the full mass range using the PDF $n_S P_S
+ n_B P_B$, where the signal and background yields $n_S$ and $n_B$ are
determined in the fit.
The statistical significance $S = {\rm sign}(n_S) \sqrt{2 \log({\cal
    L}_S/{\cal L}_B)}$, where ${\cal L}_S$ is the maximum likelihood
for $n_s$ signal events over the background yield,
and ${\cal L}_B$ is the likelihood for $n_S = 0$, is calculated for
each scan point.  The distributions of $S$ values for
all the scan points are nearly normal.

Significance values greater than $3$ are found in two scan points, both
in the \mumu mode in the $\FourS+{\rm off\mbox{-}resonance}$ sample.
The highest significance is $S=4.7$, with a signal yield of
13 events at the low-mass threshold of $m_0=0.212\gevcc$. 
The second-highest significance of $S=4.2$ occurs at $m_0=1.24\gevcc$,
corresponding to a signal yield of 10 events. 
To obtain the $p$-values for these significances, we perform the scans on a large
number of $m_{\mumu}$ spectra generated according to the background PDF, obtained from
the data with a finer binning of $n=5$. 
With this choice of $n$, the generated spectra are not sensitive to fluctuations of the order of the signal resolution (which correspond to $n=1$), yet include features that are much smaller than the resolution of the PDF ($n=15$).
We find that the probability for $S\ge 4.7$ ($4.2$) anywhere in the
\mumu spectrum with $\m_{\mumu} < 0.37\gevcc$ ($m_{\mumu} >0.5\gevcc$)
is $4\times 10^{-4}$ ($8\times 10^{-3}$).
The $p$-values are consistent with the naive expectation $p(S) w/R$, where $p(S)$ is the $p$-value without the ``look-elsewhere effect", $w$ is the width of the mass region under study, and $R$ is the average value of $R_i$. 
We do not include the other modes in the calculation of the $p$-values. Doing so would naively multiply the $p$-values by about six.
Further study provides strong indication for
material-interaction background in the $0.212\gevcc$ region. Specifically, most
of the 34 \mumu vertices with $m_{\mumu} < 0.215\gevcc$ occur inside
or at the edge of detector-material regions, including 10 of the vertices that also
pass the $e^+e^-$ selection criteria and 10 that pass the $\pi^+\pi^-$
criteria. Thus, the peak is consistent with misidentified photon
conversions and hadronic interactions close to the mass threshold. 
We conclude that a significant signal is not observed.

Systematic uncertainties on the signal yields are calculated for each
scan fit separately.
The dominant uncertainty is due to the background PDF, and is
evaluated by repeating the scans with $n=20$, which is the maximal plausible value for $n$ that does not lead to a large probability for false-signal detection. This uncertainty is a few signal events on average, and generally decreases with mass.
An additional uncertainty is evaluated by taking $R_i$ 
from events with $\pt < 0.8\gevc$ or $\pt > 0.8\gevc$.
To estimate uncertainties due to the weak signal PDF dependence on $r$
and $m$, we repeat the scans after obtaining $H^i_S$ from signal MC events
with either $r<4$ cm or $r>4$ cm, as well as from signal MC events from
adjacent mass regions.
The uncertainty due to the signal mass resolution is evaluated by
comparing the mass pull distributions of \KS mesons in data and MC, whose widths differ by 5\%.
A conservative uncertainty of 2\% on the signal reconstruction efficiencies 
is estimated from the \KS reconstruction efficiency in data and
MC.
Smaller uncertainties on the efficiency, of up to 1\%, arise from 
particle identification, and signal MC
statistics.  The total uncertainties on the efficiency are reported in
the efficiency tables~\cite{ref:supp}.

Observing that the likelihood ${\cal L}_S$ is a nearly normal
function of the signal yield, it is analytically convolved with a
Gaussian representing the systematic uncertainties in $n_S$, obtaining the
modified likelihood function ${\cal L}'_S$.  The 90\% confidence level
upper limit $U_S$ on the signal yield is calculated from $\int_0^{U_S}
{\cal L}' {\rm d}n_S/\int_0^\infty {\cal L}'{\rm d}n_S=0.9$. Dividing
$U_S$ by the luminosity yields an upper limit on the product
$\sigma(\epem\to LX)~\BR(L\to f)~\epsilon(f)$.  This limit is shown 
for each mode as
a function of $m_0$ in Fig.~\ref{fig:UL-gen}, and given 
in the supplemental material~\cite{ref:supp}.

\begin{figure}[!htbp]
\includegraphics[width=\columnwidth, bb=40 20 550 670]{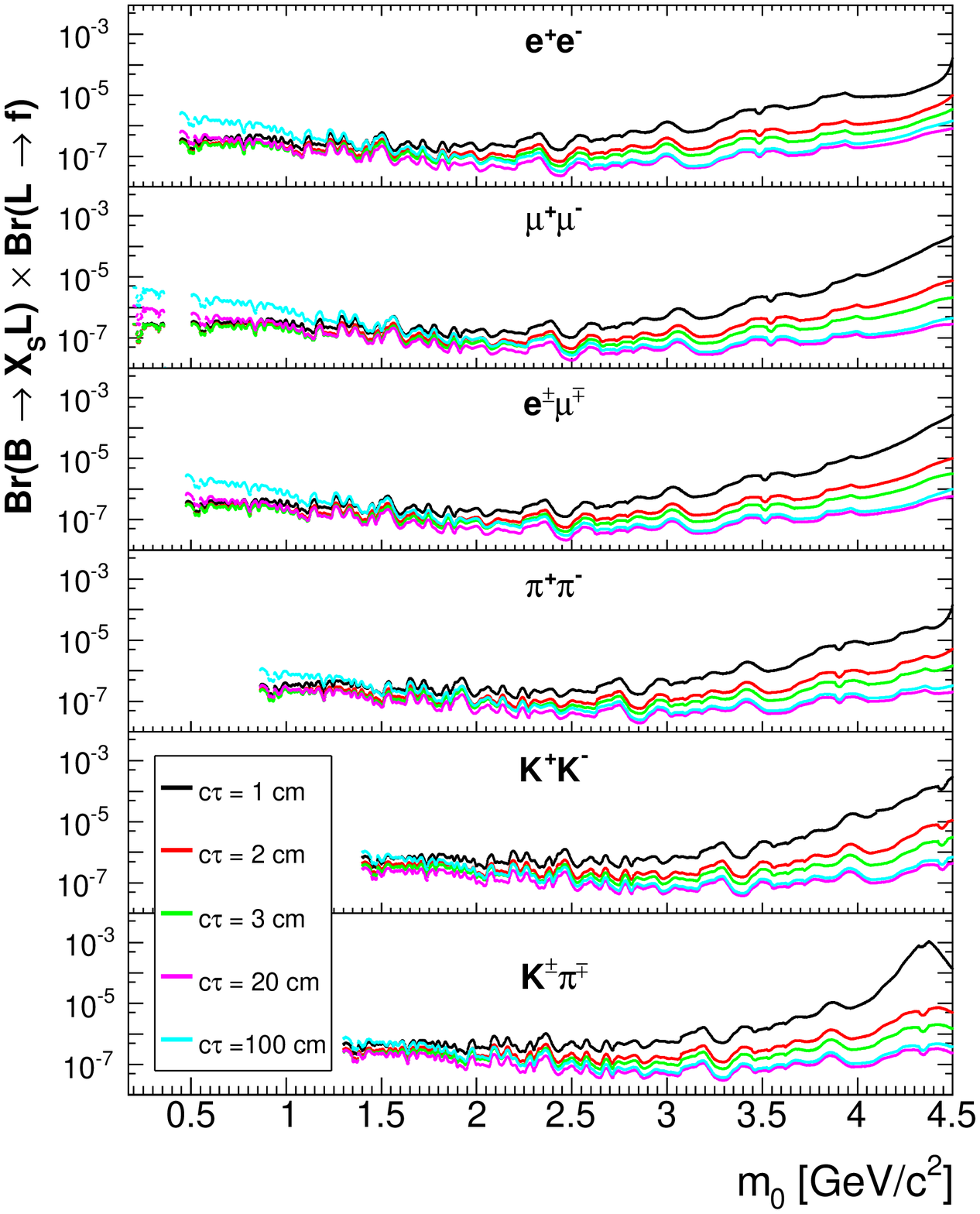}
\caption{\label{fig:UL-B} Implications of the results for Higgs-portal
  scenarios, showing the 90\% confidence level upper limits on the
  product of branching fractions $\BR(B\to X_s L) \BR(L\to f)$ as a
  function of $L$ mass for each final state $f$ and for different
  values of $c\tau$. The limits include all systematic uncertainties.
}
\end{figure}

Determining the efficiency from the $B\to X_s L$ signal MC sample, we
obtain upper limits on the product of branching fractions $\BR(B\to
X_s L) \BR(L\to f)$ for each of the final states $f$.  These limits
are shown in Fig.~\ref{fig:UL-B}.
%

\begin{figure}[!htbp]
\includegraphics[width=\columnwidth, bb=40 20 550 670]{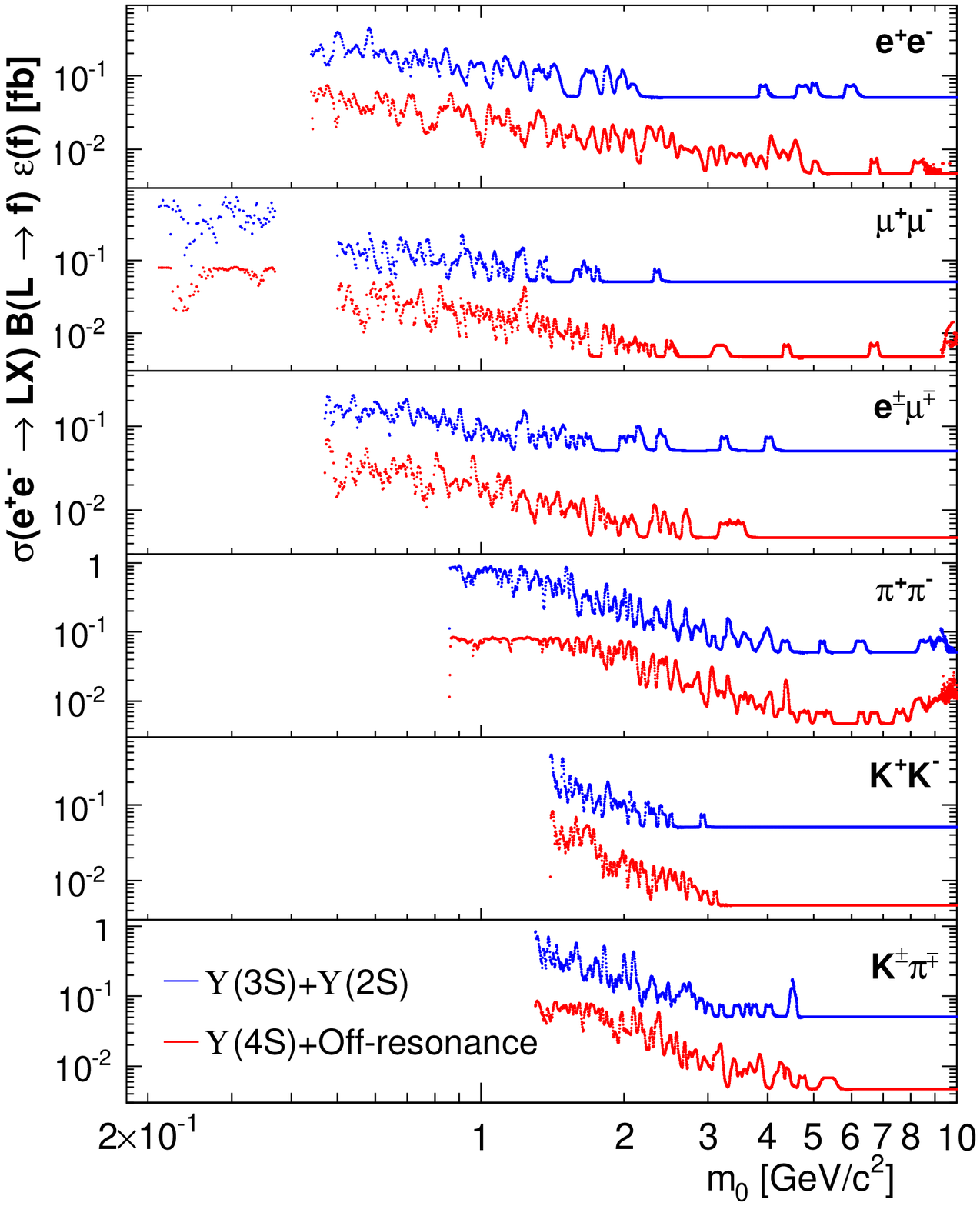}
\caption{\label{fig:UL-gen} The 90\% confidence level upper limits on
  $\sigma(\epem\to LX) \BR(L\to f) \epsilon(f)$ as a function of $L$ mass
  for the $\FourS+{\rm off\mbox{-}resonance}$ sample (red lower points) and for the 
  $\ThreeS+\TwoS$ sample (blue upper points). The limits include the
  systematic uncertainties on the signal yield. }
\end{figure}

In conclusion, we have performed a search for long-lived particles $L$
produced in $\epem$ collisions. No signal is observed, and upper
limits on $\sigma(\epem\to LX)~\BR(L\to f)~\epsilon(f)$ and on
$\BR(B\to X_s L) \BR(L\to f)$ are set at 90\% confidence level for six two-body final states
$f$. We provide detailed efficiency tables to enable application of our
results to any specific model~\cite{ref:supp}.

We are grateful for the excellent luminosity and machine conditions
provided by our \pep2\ colleagues, 
and for the substantial dedicated effort from
the computing organizations that support \babar.
The collaborating institutions wish to thank 
SLAC for its support and kind hospitality. 
This work is supported by
DOE
and NSF (USA),
NSERC (Canada),
CEA and
CNRS-IN2P3
(France),
BMBF and DFG
(Germany),
INFN (Italy),
FOM (The Netherlands),
NFR (Norway),
MES (Russia),
MINECO (Spain),
STFC (United Kingdom),
BSF (USA-Israel). 
Individuals have received support from the
Marie Curie EIF (European Union)
and the A.~P.~Sloan Foundation (USA).


\end{document}